\documentstyle[11pt,newpasp,twoside,epsf]{article}
\def\edcomment#1{\iffalse\marginpar{\raggedright\sl#1\/}\else\relax\fi}
\reversemarginpar

\begin{document}
\title{Gamma rays from clusters of galaxies}
\author{Pasquale Blasi}
\affil{INAF/Osservatorio Astrofisico di Arcetri\\
Largo Enrico Fermi, 5\\
I-50125 Firenze (ITALY)}

\begin{abstract}
The search for gamma radiation in clusters of galaxies represents a 
precious tool to investigate the history of these large scale structures. 
Clusters or sources within them accelerate cosmic rays, as demonstrated 
by the detection of radio halos, hard X-rays and UV emission, and confine 
them over cosmological time scales. 
Nonthermal and thermal phenomena may be closely related and 
observations of gamma rays may tell us about this link. In this paper we 
review the physics of cosmic ray acceleration and confinement in clusters 
of galaxies and the related gamma ray signatures. In particular we describe
in some detail the role of cluster mergers for the acceleration of nonthermal
particles. The perspectives for gamma ray detection with GLAST and with 
ground based detectors are also discussed.
\end{abstract}

\section{Introduction}

The presence of nonthermal particles in clusters of galaxies is a well 
established fact. These particles are responsible for extended synchrotron 
radio halos in several clusters (see Feretti et al., 2000 for a recent 
rewiew), as well as for hard X-ray (HXR) and extreme ultraviolet (EUV) 
excesses (see e.g. Fusco Femiano et al., 1999, 2000; Lieu et al., 1996).
Several explanations have been proposed for the origin of this radiation, but
at present there is no conclusive evidence in favor or against any of these
models. 

The simplest explanation for the HXR excess is based on the inverse compton
scattering (ICS) of the same relativistic electrons that are responsible for
the radio halos. In this case, low values of the intracluster magnetic field
are required, that seem in some contradiction with the much larger values of
the fields evaluated through faraday rotation measurements (RM) (Eilek 1999;
Clarke, Kronberg \& B\"{o}ringer 1999). 
These measurements are however quite difficult, and the discrepancy needs 
to be considered critically. The severe energy losses associated with the 
relativistic electrons make the nonthermal phenomena due to synchrotron and
ICS of relativistic electrons transient phenomena, with duration not longer
than a few billion years. However, if the acceleration processes occur during
some violent phenomenon such as cluster mergers, some level of reacceleration
may be expected due to the presence of turbulence in the intracluster medium
(Brunetti et al. 2001a, 2001b).

Alternative explanations of the HXR excess have also been proposed, based on
acceleration of electrons from the thermal bath and bremsstrahlung radiation
from these particles (Ensslin, Lieu \& Biermann 1999; Blasi 2000; Dogiel 2000;
Sarazin \& Kempner 2000). These models also have their drawbacks, as discussed
by Petrosian (these proceedings) and Petrosian (2001).

An important theoretical insight transformed our way of looking at clusters:
cosmic rays accelerated in clusters are trapped there for cosmological times
(Berezinsky, Blasi, \& Ptuskin 1997; V\"{o}lk, Aharonian, \& Breitschwerdt
1996). Clusters behave as cosmological storage rooms for cosmic rays.
The combination of this argument and the ever-increasing mass of observations
of nonthermal phenomena, generated an unprecedented interest in clusters as
possible sources of gamma rays. The detection (or not) of gamma radiation
by one of the future gamma ray telescopes such as GLAST, or even current
ground based telescopes such as STACEE or HEGRA would allow us to weigh the 
nonthermal content of clusters and achieve a better understanding of the 
nonthermal history of these large scale structures.

The issue of nonthermal radiation is clearly related to the problem of 
acceleration of  particles: although the common wisdom is that the acceleration
occurs during mergers of subclusters, there are several arguments which 
complicate this simple picture. We discuss this important issue at length 
in this review.

Throughout the paper we assume a flat cosmology ($\Omega_0=1$) 
with $\Omega_m=0.3$, $\Omega_{\Lambda}=1-\Omega_m=0.7$ and a value for the 
Hubble constant of $70\, \rm{km/s/Mpc}$.

The paper is planned as follows: in \S 2 we discuss the physics of cosmic ray 
confinement in clusters of galaxies; in \S 3 we summarize the gamma ray 
predictions for gamma rays from clusters of galaxies. \S 4 is devoted to 
the investigation of merger shocks as cosmic ray accelerators. The consequences
of gamma ray production from clusters onto the diffuse gamma ray background 
are discussed in \S 5, while our conclusions are presented in \S 6.

\section{Cosmic Ray confinement. When $\gamma$-rays became an option}
\label{sec:confine}

The bulk of high energy particles in clusters of galaxies propagate 
diffusively. The diffusion time scale for a particle with energy $E$ 
on a spatial scale $R$ comparable with the size of a cluster is 
$$ \tau_{diff} \approx \frac{R^2}{4 D(E)},$$
where $D(E)$ is the diffusion coefficient. It is easy to see that the 
energy at which the diffusion time becomes shorter than the age of the
universe is 
\begin{equation}
{\tilde E} = 2\times 10^8 B_\mu \rm{GeV}
\end{equation}
for a Bohm diffusion, and
\begin{equation}
{\tilde E} = 3\times 10^4 B_\mu L_{20}^{-2}\rm{GeV}
\end{equation}
for a Kolmogorov spectrum of the fluctuations on the magnetic field $B_\mu$
(in $\mu G$). $L_{20}$ represents here the scale where there is most of
the power in the Kolmogorov spectrum , in units of 20 kpc.
These expressions for $\tilde E$ tell us that the bulk of cosmic rays is 
confined within clusters, which therefore behave as {\it cosmological storage 
rooms} for cosmic rays (Berezinsky, Blasi \& Ptuskin 1997; V\"{o}lk, 
Aharonian, \& Breitschwerdt 1996). Any process or any source that accelerates 
particles within the cluster volume contributes to increase the nonthermal 
content of the intracluster gas. The present value of the energy density of 
cosmic rays in a cluster is the result of all these processes integrated over 
the lifetime of the cluster (comparable to the age of the universe $t_0$).

This argument is of special importance for particles whose energy losses occur
on time scales longer than $t_0$, in particular high energy protons (or 
nuclei). 
Relativistic electrons with $\gamma>300$ lose energy on time scales shorter 
than $t_0$, so that their energy is radiated away through synchrotron and 
ICS emission. 
The fate of these high energy electrons is to finally pile up at lorentz 
factors around $\sim 100$ where the timescale for losses, dominated 
now by Coulomb scattering, becomes of several billion years. 

For protons, the main channel of energy losses is provided by inelastic 
proton-proton scattering, with inclusive cross section 
$\sigma_{pp}\sim 3\times 10^{-26}{\rm cm^2}$. In a cluster of galaxies, the 
timescale associated with this process is 
$$\tau_{pp} = \frac{1}{n_{gas} \sigma_{pp} c} = 3.5\times 10^{10} n_{-3}^{-1}
~{\rm yrs},$$
where $n_{-3}=n_{gas}/10^{-3}{\rm cm^{-3}}$ and $n_{gas}$ is the gas density 
in the intracluster medium. Inelastic $pp$ scattering is weak enough to 
allow the accumulation of protons over cosmological times, as discussed above,
but also efficient enough for the continuous production of pions, which in 
turn decay into gamma rays (for neutral pions), electron-positron pairs and 
neutrinos (for charged pions). The decay chain is as follows:
$$p+p \to \pi^0 + \pi^+ + \pi^- + \rm{anything}$$
$$\pi^0 \to \gamma \gamma$$
$$\pi^\pm \to \mu + \nu_\mu ~~~ \mu^\pm\to e^\pm \nu_\mu \nu_e.$$
The role of electron-positron pairs from 
$\pi^\pm$ decays is subject of much debate and investigation (see also talk 
by Brunetti, these proceedings). In fact synchrotron emission from
these pairs may well reproduce the general features of the radio halos and
their diffuse appearance (Colafrancesco \& Blasi 1998), without invoking any
additional reacceleration processes. Reacceleration is instead required in 
radio halo models based on the acceleration of primary electrons and their 
propagation in the cluster volume. In order to accomodate the HXR emission 
observed from the Coma cluster, this model requires a magnetic field of  
$\sim 0.1 \mu G$. This conclusion actually holds for any other model, with
the possible exception of models in which a cutoff in the electron spectrum
is tuned up in order to reduce the corresponding synchrotron emission. In
these cases the magnetic field can be as high as $0.3-0.4\mu G$ (Brunetti et
al. 2001a). 
It was shown by Colafrancesco \& Blasi (1998) that for the Coma cluster, 
in the context of the secondary electron model, small magnetic fields imply an 
overproduction of gamma radiation compared to the EGRET upper limit 
(Sreekumar et al. 1996). 
This conclusion may possibly be avoided only if the emission regions of HXRs 
and radio radiation are different. 
A careful investigation of all these effects is being carried out by 
Blasi, Brunetti \& Gabici (2002), in order to understand under which 
conditions, if any, the fine structure of radio halos (spatial distribution 
of the radiation, spectral steepening in the outer regions of clusters, 
radio halo statistics) can be accomodated within the so-called secondary 
electron model (Dennison 1980, Colafrancesco \& Blasi 1998, Blasi \& 
Colafrancesco 1999), where all the electrons responsible for the radio halo 
are due to $pp$ scatterings.
This investigation is extremely important even if the bulk of the observed
nonthermal radiation had to be generated mainly by something other than the
secondary $e^+e^-$ pairs. In fact, if to take our own Galaxy as a template
of cosmic ray behaviour, it seems likely that in clusters, as well as in the
Galaxy, protons outnumber electrons (at least around 1 GeV) by about a factor
100. The questions then become: where are these protons? And how can
we detect them? The answer to these questions is, we believe, in gamma ray 
observations, both in the GeV range, with GLAST, and in the TeV range, with
ground based detectors. In the next section we describe the current predictions
of gamma ray fluxes and their physical information load.

\section{$\gamma$-ray emission}

A useful way of discussing the gamma ray emission from clusters of galaxies 
is by simply parametrizing the proton abundance in clusters as a fraction of 
the thermal (virial) energy of the cluster, or, in other words, in terms of 
deviations from equipartition.
In fig. 1, we plot the gamma ray fluxes expected from a Coma-like cluster
as due to $pp$ inelastic scattering, and neglecting at this stage any other 
contribution. The energy density of cosmic ray protons is taken to be equal
to the thermal energy density.
\begin{figure}
\plotfiddle{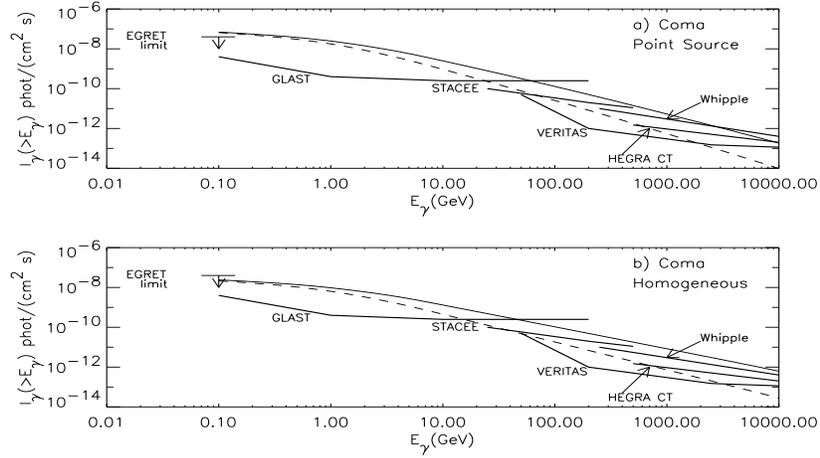}{6cm}{0}{60}{50}{-180}{-180}
\caption{Predicted gamma ray flux from the Coma cluster as produced by
$\pi^0$ decay (Blasi 1999). 
The injection spectrum of protons is $E^{-2.1}$ for the
solid line and $E^{-2.4}$ for the dashed line. In both cases cosmic rays
are assumed in equipartition with the thermal energy. The sensitivities of some
present and future gamma ray experiments are also plotted. The injection 
occurs from a point source in the cluster's center (upper panel) or 
homogeneously (lower panel).}
\end{figure}
In the upper panel we assumed that cosmic rays are injected by a point source
(for instance a radio Galaxy) in the center of the cluster, while in the 
bottom panel cosmic rays are assumed to be injected homogeneously in the 
cluster volume. The EGRET upper limit and the sensitivity curves for GLAST 
and for some ground based experiments are also reported in the figure. 
Some comments are in order: 1) the fluxes of gamma rays above $100$ GeV are
already at the EGRET sensitivity level, and will certainly be at hand for 
GLAST. 2) In the presence of protons, it is basically unavoidable to have 
gamma ray production up to at least the TeV range; in this energy region, 
the radiation spectra reproduce the spectrum of parent protons.

It seems that even with present experiments, such as STACEE, it would be 
possible to put interesting upper limits on the gamma ray fluxes from some
nearby clusters of galaxies, such as Coma (Blasi 1999). If the clusters are
too far away, then absorption effects due to photon-photon pair production
on the cosmic infrared background and smaller fluences may make the detection 
more difficult or even impossible at energies in excess of a few TeV.

In order to have a more complete picture of the processes that contribute to
the gamma ray brightness of a cluster, we need to include at least two other
components, namely primary and secondary electrons. Blasi (2000) has carried
out this calculation for the case of a merger shock as accelerator, but the
basic features remain valid even for a different type of accelerator. A
similar numerical calculation was carried out by Miniati et al. (2001).

For a Coma-like cluster, the flux of gamma radiation is plotted in fig. 2.
\begin{figure}
\plotfiddle{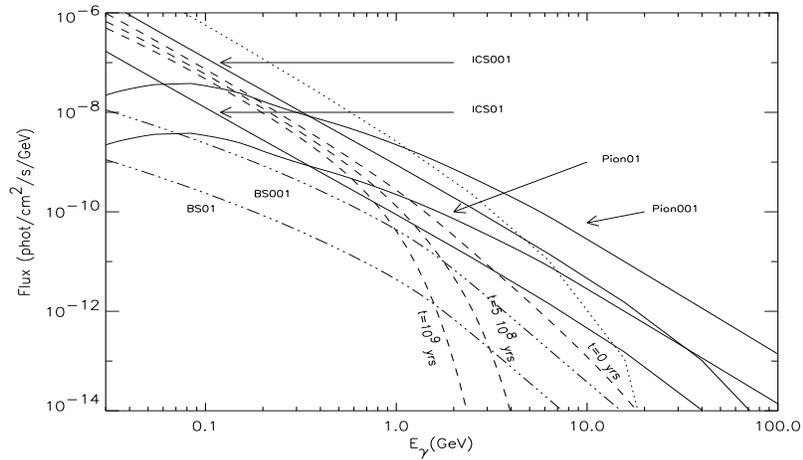}{6cm}{0}{60}{50}{-180}{-180}
\caption{Predicted gamma ray emission from the Coma cluster (Blasi 2000). 
A detailed description of the curves can be found in the text.}
\end{figure}
The spectrum of the injected particles has been chosen to reproduce the 
spectral slope of the radio halo in Coma. The solid thick lines, labelled as 
{\it Pion01} and {\it Pion001} are the gamma ray fluxes due to neutral pion
decay in $pp$ inelastic scattering. The normalization between electrons and
protons accelerated at the same merger shock (therefore with the same power
law injection spectrum in momentum) is parametrized by a 
parameter $\xi$ representing the ratio in the number density of electrons 
and protons at injection. After fixing $\xi$ the solution of 
the transport equation (including diffusion and energy losses) can be
determined.
For the numerical calculations we use $\xi=0.1$ (curve {\it Pion01}) 
and $\xi=0.01$ (curve {\it Pion001}).
The thin solid lines labelled as {\it ICS01} and {\it ICS001} represent the
gamma ray fluxes due to ICS of the secondary electrons from the decay of 
charged pions for the two values of the parameter $\xi$. The two curves 
labelled {\it BS01} and {\it BS001} are the gamma ray fluxes due to 
bremsstrahlung emission of secondary electrons. The dashed lines are the 
result of bremsstrahlung emission of primary electrons, as a function of the
time from the end of the merger event. 

The dotted line is the flux of gamma rays due to ICS of the primary electrons.
Note that this conctribution exists only for Bohm diffusion coefficient, as the
acceleration time of electrons exceeds the timescale for losses for less
extreme choices of the diffusion coefficient. In these more reasonable cases,
the ICS emission ends up in the X-rays and stops there. 

Some very general conclusions can be extracted from fig. 1:
\begin{itemize}
\item[{\it i)}] The fluxes that are expected from a Coma-like cluster are 
accessible to GLAST in the energy region above 100 MeV. 
\item[{\it ii)}] While the gamma ray emission from primary electrons is always
time dependent and rapidly fading away after the end of the merger (it
basically disappears a few hundred million years after the merger), the gamma 
ray emission due to either neutral pion decay or emission of secondary 
electrons is \underline{time independent} (and actually slightly increasing
with time), as a result of cosmic ray confinement and of slow energy losses of 
relativistic protons. 
\item[{\it iii)}] The contribution due to protons is underestimated in fig. 2,
because only one merger has been considered. While this is an excellent 
approximation for relativistic electrons, it is not for protons, which are
accumulated during the cluster's history piling up to a likely larger value
than that used to obtain the curves in fig. 2. Their spectrum depends however
on the specific injection mechanism.
\item[{\it iv)}] Above a few GeV the gamma ray flux is likely to be dominated
by the decay of neutral pions. The spectra are as flat as those of the parent
protons and may extend to very high energies. In the energy region above a
few hundred GeV these fluxes will be accessible to ground based experiments 
(see also fig. 1) such as VERITAS, MAGIC and HESS.
\item[{\it v)}] The ICS contribution of primary electrons is present only if
electrons can attain a maximum Lorentz factor larger than about $10^6-10^7$,
which may happen only for very small diffusion coefficients, like in the Bohm 
case (Blasi 2000). These considerations are crucial for models that try
to establish a connection between cluster mergers and the diffuse extragalctic
gamma ray background (Loeb \& Waxman 2000).

\end{itemize}

The main conclusions listed above remain valid even if
the nonthermal particles are accelerated at a place other than merger shocks.
Nevertheless, much interest has been shown recently on merger events as the 
origin of nonthermal phenomena in clusters of galaxies. Therefore in the next
section, we discuss in detail the acceleration of particles in merger shocks,
and their relevance on cosmological time scales.

\section{Merger shocks as cosmic ray accelerators}

Relativistic particles can be accelerated at strong shocks by diffusive
(first order) Fermi acceleration (Fermi 1949; Blandford \& Eichler 1987).
This mechanism has been invoked several times as the ideal acceleration 
process in clusters of galaxies that have been involved in a merger
event (Blasi 2000; Fujita \& Sarazin 2001). In \S 4.1 we briefly summarize 
the basic physics of shock acceleration, since it is 
instrumental to understand whether merger related shock waves can indeed 
play a role for the acceleration of the relativistic particles responsible 
for the observed nonthermal radiation from clusters of galaxies. 
In order to assess this point, we also need to reconstruct the merger history 
of a cluster, and use it to determine the statistics of strengths of the 
shocks associated to the merger events. 
We do this in \S 4.2. The results summarized here are discussed at length 
by Gabici \& Blasi (2002).

\subsection{The basics of shock acceleration in clusters \label{sec:shocks}}

A shock with compression factor $r$ and Mach number ${\cal M}$ can accelerate
particles to a power law in momentum $f(p)\propto p^{-\alpha}$, with slope
$\alpha$ related to the Mach number and compression factor by the following
expressions:
\begin{equation} 
\alpha=\frac{r+2}{r-1} = 2 \frac{{\cal M}^2 + 1}{{\cal M}^2 - 1}.
\label{eq:slope}
\end{equation}
The acceleration occurs diffusively, in that particles scatter back and
forth the shock, gaining at each crossing and recrossing an amount of
energy proportional to the energy of the particle itself, 
$\Delta E/E\sim V/c$, where $V$ is the speed of the shock and $c$ is the
speed of light.
The distribution function of the accelerated particles is normalized here 
by $\int_{p_{min}}^{p_{max}} dp E(p) f(p) = \eta \rho u^2$, where $E(p)=
\sqrt{p^2 + m^2}$ and $m$ is the mass of the accelerated particles, $\eta$
is an efficiency of acceleration, $\rho$ and $u$ are the density and speed
respectively of the fluid crossing the shock surface. The minimum and maximum
momenta ($p_{min}$ and $p_{max}$) of the accelerated particles are determined 
by the properties of the shock. In particular, $p_{max}$ is the result of the
balance between the acceleration rate and either the energy loss rate or
the rate of escape from the acceleration region. Less clear is how to 
evaluate $p_{min}$; the minimum momentum of the particles involved in the
acceleration process depends on the microphysics of the shock, a problem
well known in the investigation of shock acceleration as the 
{\it injection problem}. Fortunately, most physical observables usually
depend very weakly on $p_{min}$.  

In the following we estimate the value of the maximum energies for electrons
and protons as accelerated particles. The acceleration time, as a function of 
the particle energy $E$ can be written as
\begin{equation}
\tau_{acc} (E) = \frac{3}{u_1-u_2} D(E) \left[\frac{1}{u_1}+\frac{1}{u_2}
\right] = \frac{3 D(E)}{u_1^2} \frac{r(r+1)}{r-1},
\end{equation}
valid for any choice of the diffusion coefficient $D(E)$, for which we
consider two possible models. First we use the expression proposed in 
(Blasi \& Colafrancesco 1999): 
\begin{equation} 
D(E) = 2.3\times 10^{29} B_{\mu}^{-1/3} L_{20}^{2/3} E(GeV)^{1/3} cm^2/s,
\label{eq:dif1}
\end{equation}
where $B_{\mu}$ is the magnetic field in microgauss.
Here we assumed that the magnetic field is described by a Kolmogorov power 
spectrum.

In this case the acceleration time becomes:
\begin{equation}
\tau_{acc} (E) \approx 6.9\times 10^{13} B_{\mu}^{-1/3} L_{20}^{2/3} 
E(GeV)^{1/3} v_8^{-2} g(r)~~~ s,
\end{equation}
where $v_8=\frac{v}{10^8 cm/s}$ and $g(r)=r(r+1)/(r-1)$ and $v=u_1$.

For electrons, if the average magnetic field is less than $\sim 3\mu G$, the
energy losses are dominated by ICS off the microwave background, with a loss
time $\tau_{loss}\approx 4\times 10^{16}/E~s$, where $E$ is in GeV. The maximum
energy of accelerated electrons is obtained requiring $\tau_{acc}<\tau_{loss}$:
\begin{equation}
E_{max}^e \approx 118 L_{20}^{-1/2} B_{\mu}^{1/4}  
v_8^{3/2} g(r)^{-3/4}~ GeV.
\end{equation}
The compression ratio $r$ and the velocity $v_8$ are not independent, since 
\begin{equation}
r=\frac{\frac{8}{3} {\cal M}^2}{\frac{2}{3} {\cal M}^2 + 2},
\end{equation}
valid for an ideal monoatomic gas. 

For protons, energy losses are not relevant and the maximum energy is clearly
determined by the finite time duration of the merger event. Therefore the 
maximum energy for protons will be defined by the condition $\tau_{acc}<
t_{merger}$, which gives 
\begin{equation}
E_{max}^p \approx 9\times 10^7 L_{20}^{-2} B_{\mu} v_8^6 g(r)^{-1/2}~GeV.  
\end{equation}

As a second possibility for the diffusion coefficient we assume Bohm diffusion,
well motivated for the case of strong turbulence. In this case:
\begin{equation}
D(E)=3.3\times 10^{22} E(GeV)/B_\mu ~ cm^2/s.
\label{eq:dif2}
\end{equation}
In this case, for electrons we obtain:
\begin{equation}
E_{max}^e \approx 6.3\times 10^4 B_{\mu}^{1/2} v_8 g(r)^{-1/2}~ GeV,
\label{eq:Emaxe}
\end{equation}
while for protons
\begin{equation}
E_{max}^p \approx 3\times 10^9 B_{\mu} v_8^2 g(r)^{-1}~GeV.  
\end{equation}
If $E_{max}^p$ becomes larger than $\sim 10^{10}$ GeV energy losses due
to pair production and photopion production on the photons of the microwave
background become important and limit the maximum energy to less that 
a few $10^{10}$ GeV. 

\subsection{The merger tree and related shocks \label{sec:history}}
The standard theory of structure formation predicts that larger structures 
are the result of the mergers of smaller structures: this hierarchical model
of structure formation in the universe has been tested in several independent
ways and provides a good description of the observations of the mass function 
of clusters of galaxies and their properties.

While a complete understanding of the process of structure formation can only 
be achieved by numerical N-body simulations, an efficient and analytical 
description can also be obtained and several of these approaches are widely 
discussed in the literaure. Historically, the first approach to the problem 
was proposed by Press \& Schechter (1974, hereafter PS) and successively 
developed by Bond et al. (1991) and Lacey \& Cole (1993, hereafter LC) among 
others. 
In the PS formalism, the differential comoving number density of clusters with 
mass $M$ at cosmic time $t$ can be written as:
\begin{equation}
\frac{dn(M,t)}{dM}=\sqrt{\frac{2}{\pi}}\,\frac{\varrho}{M^2}\,
\frac{\delta_c(t)}
{\sigma(M)}\,\left|{\frac{dln \sigma(M)}{dln M}}\right| exp\left[-\frac
{\delta_c^2(t)}{2\sigma^2(M)}\right].
\end{equation}
The rate at which clusters of mass $M$ merge at a given time $t$ is written 
as a function of $t$ and of the final mass $M^{\prime}$ (LC, 1993):

$$R(M,M^{\prime},t)dM^{\prime}=$$
\begin{eqnarray}
\sqrt{\frac{2}{\pi}}\,\left|
\frac{d\delta_c(t)}{dt}\right|\,\frac{1}{\sigma^2(M^{\prime})}\,
\left|\frac{d\sigma(M^{\prime})}{dM^{\prime}}\right|\,
\left(1-\frac{\sigma^2(M^{\prime})}{\sigma^2(M)}\right)^{-3/2} \nonumber \\
\rm{exp}\left[-\frac{\delta_c^2(t)}{2}\left(\frac{1}{\sigma^2(M^{\prime})}-
\frac{1}{\sigma^2(M)}\right)\right]dM^{\prime},
\end{eqnarray}
where $\varrho$ is the present mean density of the universe, $\delta_c(t)$ 
is the critical density contrast linearly extrapolated to the present time 
for a region that collapses at time $t$, and $\sigma(M)$ is the current rms 
density fluctuation smoothed over the mass scale $M$.
For $\sigma(M)$ we use an approximate formula proposed by Kitayama (1997), 
normalized by assuming a bias parameter $b=0.9$. We adopt the expression of
$\delta_c(t)$ given by Nakamura \& Suto (1997). In this respect our approach
is similar to that adopted by Fujita \& Sarazin(2001).

Salvador-Sol\'e, Solanes \& Manrique (1998) modified the model illustrated
above, by introducing a new parameter,  $\Delta_m = r_{crit}= 
[(M^{\prime}-M)/M]_{crit}$, defined as a peculiar value of the captured
mass that separates the accretion events from merger events. 
Events in which a cluster of mass $M$ captures a dark matter halo with 
mass smaller then $\Delta_m M$ are considered as continuous mass 
accretion, while events where the collected mass is larger than $\Delta_m M$ 
are defined as mergers. The value of $\Delta_m M$ is somehow arbitrary.

Using this effective description of the merger and accretion events, it
is easy to construct simulated merger trees for a cluster with fixed mass 
at the present time. 
Although useful from a computational point of view, this difference does 
not correspond to any new phyics information, therefore in the following 
we will adopt the name ``merger'' for both regimes, provided there is no 
ambiguity or risk of confusion.
\begin{figure}
\plotfiddle{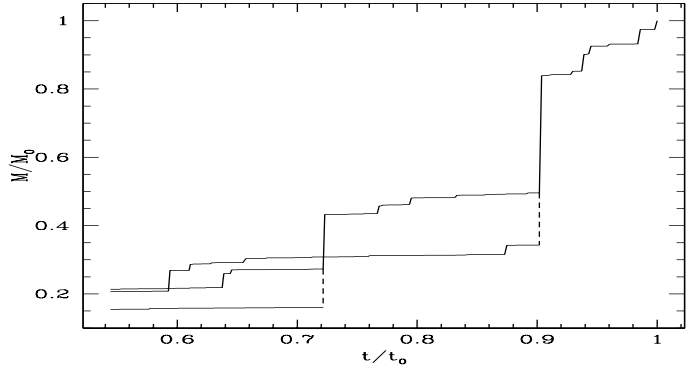}{4cm}{0}{50}{27}{-150}{-50}
\caption{Merger history of a cluster with present mass $10^{15}$ solar masses
(Gabici \& Blasi 2002). The mass (y-axis) suffers major jumps in big merger
events. Time is on the x-axis.}
\end{figure}
In figure 3 we plotted the merger tree for a cluster with 
present mass equal to $10^{15} M_\odot$ and for $\Delta_m=0.6$. The
big jumps in the cluster mass correspond to merger events, while smaller jumps
correspond to what Salvador-Sol\'e et al. (1998) defined as accretion events. 

While the dark matter components of two clusters involved in a merger can 
compenetrate each other due to the collisionless nature of dark matter,
and form a deeper gravity potential well, the baryon components of the 
clusters are forced to move supersonically during the merger event, and 
shock surfaces are formed. These shocks are instrumental for the heating 
of the intracluster gas, since they allow the conversion of part of the 
available gravitational energy into thermal energy.

In this section we describe in more detail the physical properties of such 
shocks, with special attention for their Mach numbers and compression 
factors.

We assume to have two clusters, as completely virialized structures, at 
temperatures $T_1$ and $T_2$, and with masses $M_1$ and $M_2$ (here the
masses are the total masses, dominated by the dark matter component). 
The virial radius of a cluster can be written as follows
$$
r_{vir,i} = \left(\frac{3 M_i}{4 \pi \Delta_c \rho_{cr,0} (1+z_{f,i})^3}
\right)^{\frac{1}{3}}=
$$
\begin{equation}
\left(\frac{G M_i}{100 \Omega_m H_0^2 (1+z_{f,i})^3}\right)^{\frac{1}{3}},
\label{eq:vir}
\end{equation}
where $i=1,2$, $\rho_{cr,0}=\Omega_m 1.88 10^{-29} h^2 \rm{g}~\rm{cm}^{-2}$ 
is the current value of the critical mass density of the universe, 
$z_{f,i}$ is the redshift of formation of the cluster $i$, $\Delta_c=200$ is 
the density constrast for the formation of the cluster and $\Omega_m$ is the 
matter density fraction. 
In the right hand side of the equation we used the fact 
that $\rho_{cr,0}=3 H_0^2/8\pi G$, where $H_0$ is the Hubble constant.
The formation redshift $z_f$ is on average a decreasing function of the mass,
meaning that smaller clusters are formed at larger redshifts, consistently
with the hierarchical scenario of structure formation. There are intrinsic
fluctuations in the value of $z_f$ from cluster to cluster at fixed mass, 
due to the stochastic nature of the merger tree.

Two clusters with masses $M_1$ and $M_2$ collide with a relative velocity 
$V_r$ that can be easily calculated from energy conservation:
\begin{equation}
-\frac{G M_1 M_2}{r_{vir,1}+r_{vir,2}} + \frac{1}{2} M_r V_r^2 = 
-\frac{G M_1 M_2}{2 R_{12}},
\end{equation}
where $M_r=M_1 M_2 / (M_1+M_2)$ is the reduced mass and $R_{12}$ the 
turnaround radius of the system of two masses. In a Einstein-De Sitter
cosmology, the latter equals twice the virial radius of the system, so
that, using eq. 15, we get:
\begin{equation}
R_{12} = \left( \frac{M_1+M_2}{M_1}\right)^{1/3} r_{vir,1}.
\end{equation}
In different cosmologies this expression still remains valid in approximate 
way. The sound speed of the cluster $i$ is given by 
$$
c_{s,i} = \gamma_g (\gamma_g - 1) \frac{G M_i}{2 r_{vir,i}}
$$
where we used the virial theorem to relate the gas temperature to the 
mass and virial radius of the cluster. The adiabatic index of the gas is
$\gamma_g=5/3$. Following Takizawa (1999), the Mach numbers of each cluster 
while moving in the volume of the other cluster can be written as:
\begin{eqnarray}
{\cal{M}}^2_1 & = & \frac{4(1+\eta)}{\gamma(\gamma-1)}
\left[\frac{1}{1+\frac{1+z_{f,1}}
{1+z_{f,2}}\eta^{1/3}}-\frac{1}{4\frac{1+z_{f,1}}{1+z_{f}}(1+\eta)^{1/3}}
\right] \nonumber \\
{\cal{M}}^2_2 & = & \eta^{-2/3}\frac{1+z_{f,1}}{1+z_{f,2}}{\cal{M}}^2_1, 
\end{eqnarray}
where $\eta=M_2/M_1<1$ and $z_f$ is the formation redshift of the cluster
with mass $M_1+M_2$. 
Our strategy at this point is to consider a cluster with mass $M_0$ at the
present time and simulate numerous merger trees and calculate the Mach numbers
of the subclusters taking part to the merger events. To start with, we 
simulate 500 realizations of the merger history of a $10^{15} M_\odot$ 
cluster. A value $\Delta_m=0.05$ is assumed, much lower than in 
(Fujita \& Sarazin 2001). This simply implies that we follow the histories of 
very small halos of dark matter, rather than the big ones only. 
The results of our calculations of the Mach numbers are plotted in fig. 4a 
(left panel). 
\begin{figure}
\plottwo{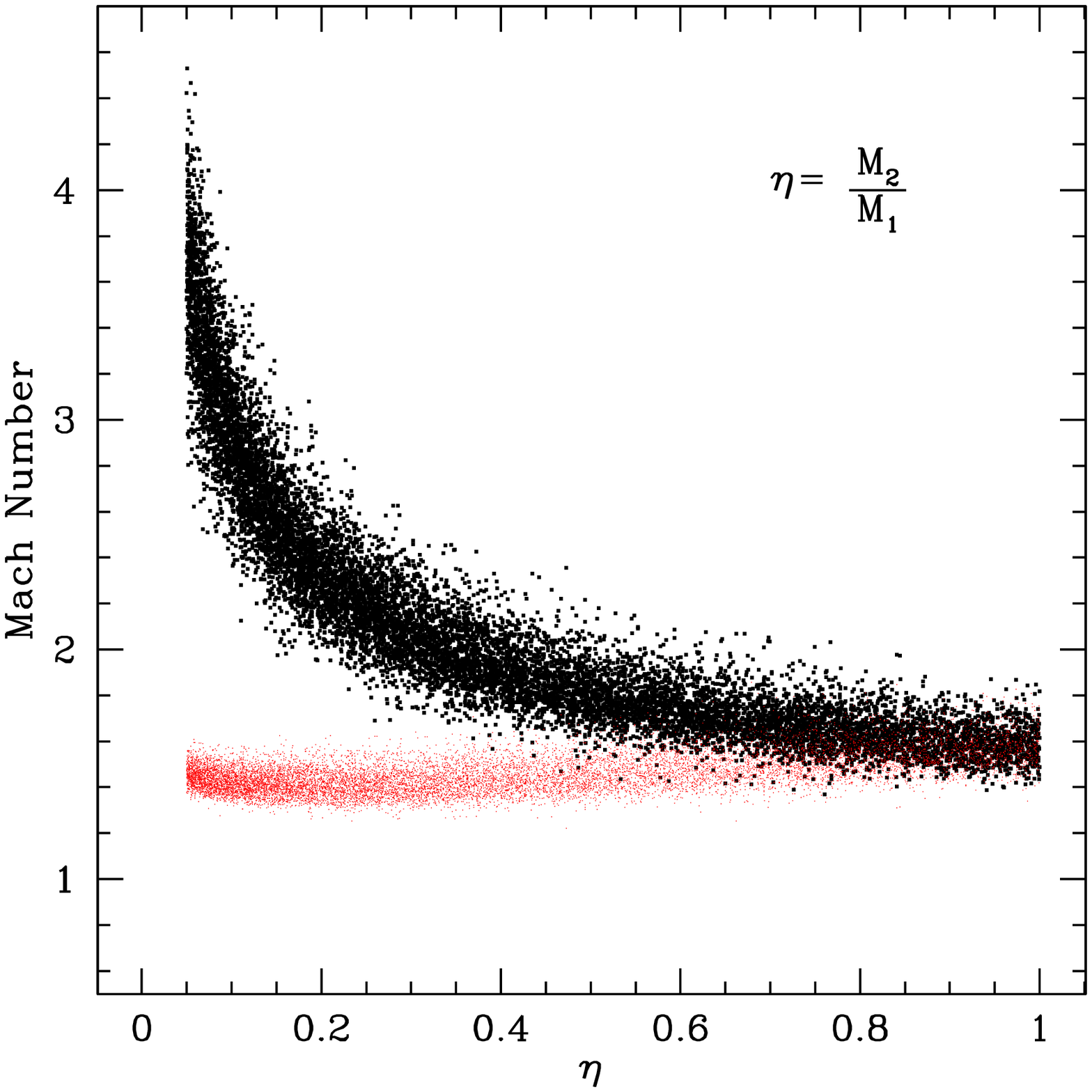}{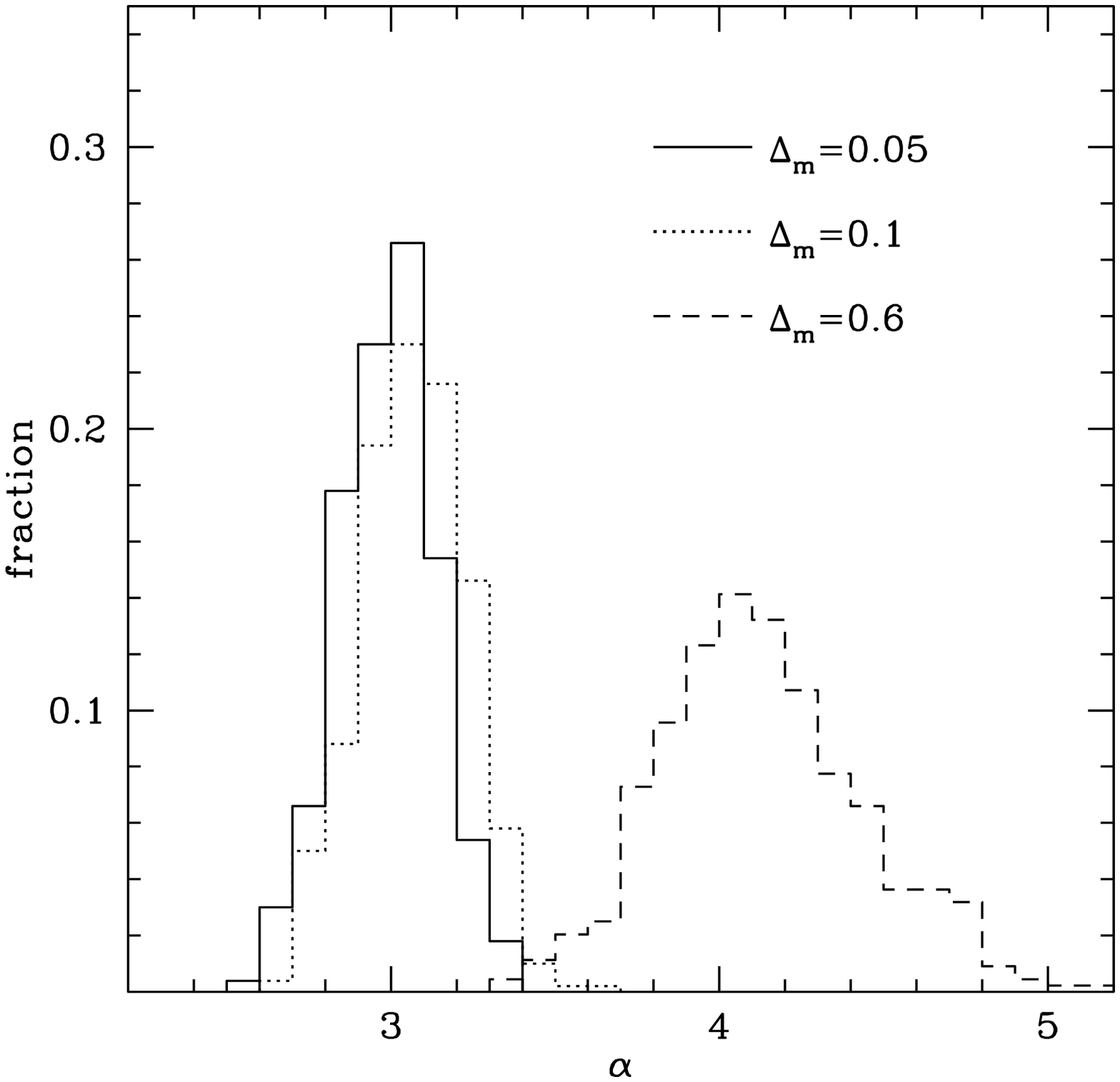}
\caption{{\it 4a (left panel)}: Distribution of the Mach numbers of 
merger related shocks as a function of the mass ratio of the merging
subclusters. The upper strip is the distribution of Mach numbers in
the smaller cluster, while the lower strip refers to the bigger cluster.
{\it 4b (right panel):} Slope of the time-integrated proton
spectrum resulting from all the mergers in a cluster. The three curves
are obtained for $\Delta_m=0.05$ (solid line), $\Delta_m=0.1$ (dotted line)
and $\Delta_m=0.6$ (dashed line).}
\end{figure}
It is evident from the figure that for major mergers, involving clusters 
with comparable masses ($\eta\sim 1$), the Mach numbers of the shocks are 
of order unity. In other words the shocks are only moderately supersonic. 
In order to achieve Mach numbers of order of $3-4$ it is needed to consider 
mergers between clusters with very different masses ($\eta\sim 0.05$), which, 
in the language of Salvador-Sol\'e et al. (1998) and Fujita \& Sarazin (2001) 
would not be considered as mergers but rather as continuous accretion. These
events are the only ones that produce strong shocks, and this is of crucial 
importance for the acceleration of suprathermal particles, as discussed 
below. This picture seems to be confirmed by X-ray observations (see for 
instance Markevitch, Sarazin and Vikhlinin (1999)). 

Fig. 4b (right panel) requires some preliminary comments: during a merger 
event, new particles are accelerated from the thermal pool, but the protons 
already confined in the cluster volume are also reaccelerated. These processes
are taken into account in detail by Gabici \& Blasi (2002). 
The final spectra of protons at the present time, as a result of the 
superposition of all the merger events that a cluster suffered, have a 
shape which can be something more complicated than a simple power law 
(Gabici \& Blasi, 2002), therefore we choose here to plot, in fig. 4b, the 
slope at fixed energy, say 10 GeV (one should remember that this corresponds 
to secondary electrons with typical energy of a few GeV). The result of our 
calculations show that the time integrated proton spectra are typically steep, 
or at least steeper than needed to explain the observed nonthermal radiation, 
even in the cases $\Delta_m\ll 1$.

The situation is slightly different for primary electrons. High energy 
electrons must be relatively young to generate appreciable nonthermal 
radiation. A typical time for the production of these electrons is of 
about one billion years. In other words, only electrons injected in the
last few mergers (or accretion events) can generate nonthermal radiation
at present. Hence, we generated a merger tree of a cluster with Coma-like
mass and extended it only for one billion years in the past, and again
calculated the spectra of the electrons accelerated at the merger shocks.
We repeated this procedure for 500 clusters and only about $30\%$ of them
suffered any kind of merger (down to $\eta=0.05$) in the last billion years.
Of these, $\sim 20\%$ are characterized by strong shocks, able to accelerate
particles with spectra flatter than $E^{-2.4}$. In other words, only 
$\sim 6\%$ of clusters with mass comparable with the Coma cluster should
have a similar radio halo. It is important to stress again that strong
shocks are not associated to major mergers but rather to 
{\it accretion events}.

\section{The extragalactic diffuse gamma ray background} 

Since the detection of an isotropic excess in the gamma ray emission of 
the Galaxy, as detected by EGRET (Sreekumar et al. 1998), and interpreted 
as of extragalactic origin, several attempts have been made of relating it to
nonthermal processes occurring in clusters of galaxies. The initial paper
by Dar \& Shaviv (1995) reached incorrect conclusions, mainly due to an 
errouneous calculation of the spectrum of the radiation, as later recognized
by Berezinsky, Blasi \& Ptuskin (1997). More recently, Loeb \& Waxman (2000)
have reproposed this connection: in their paper the gamma ray emission is
generated by ICS of relativistic electrons accelerated at the shocks generated
during structure formation. This scenario has received much attention and
also inspired some searches for associations between the EGRET unidentified
sources and the positions of some clusters of galaxies (Colafrancesco 2002;
Kawasaki \& Totani 2001, Totani \& Kitayama 2000). 

Some of the shocks related to the formation of large scale structures 
are related to cluster mergers, in that they form within the virial radii 
of the merging clusters.
Other shocks form in the outer regions and propagate in a colder medium,
therefore reaching higher Mach numbers (Miniati, F., et al. 2000). 
The Mach numbers of the shocks related to structure formation range between 
unity and a few hundreds. The weaker shocks, usually associated to major 
mergers, as discussed by Gabici \& Blasi (2002) are inefficient particle 
accelerators (see also the discussion in \S 4.2).
The stronger shocks extend over several Mpc regions and are strong enough 
to generate flat spectra. To understand whether the former or the latter 
dominate, it is needed to run careful numerical simulations, as performed 
by Miniati (2002). One point that should be made clear is that the possibility
that part of the diffuse extragalactic gamma ray background may have a 
connection to large scale structures relies upon the strong assumption that 
the diffusion coefficient of the particles around the shock surface is well 
described by a Bohm diffusion coefficient. 
Only in this case the electron energies may be large enough to 
generate gamma rays by ICS off the photons of the cosmic microwave backgound. 
Moreover, the magnetic fields that have been estimated from the observations 
of nonthermal radiation in clusters are of fraction of $\mu G$, but they 
only refer to the virialized region of the clusters. In the flux freezing 
approximation, the magnetic field scales with the radial coordinate as 
$B\sim \rho^{2/3}\sim r^{-3\beta}$, where $\beta\approx 0.75$ is the 
parameter entering the $\beta$-model for the density $\rho$ (simulations 
performed by Dolag, Bartelmann \& Lesch (2002) show even steeper trends). 
At the distance of the large Mach number shocks, say comparable to the 
turnaround radius of a cluster, the strength of the field is likely reduced 
to $nG$ values. For a typical scale of 5 Mpc, the maximum energy of 
accelerated electrons, as derived in eq. 11, is reduced to values that are 
barely sufficient for the production of gamma rays up to $\sim 10-30$ GeV, 
even in the case of Bohm diffusion. 
Although there are several aspects that deserve further
investigation, the possibility that at least a fraction of the alleged 
extragalactic gamma ray background may be due to processes related to 
structure formation is certainly interesting and will continue to fuel 
much interest in the years to come.

\section{Conclusions}

The diffuse medium in the intracluster volume is filled with a nonthermal 
gas of particles that are the relics of all the events occurred within the 
cluster itself. Gamma ray astronomy is an important tool to study this 
component and infer information about particle acceleration and confinement
and about the specific processes (mergers, active galactic phases, and many
others) that inject nonthermal particles in a cluster and possibly contribute 
to its heating. 

We discussed here the expectations for gamma ray fluxes in the presence of 
both protons and primary electrons in clusters. Even if the amount of hadronic
cosmic rays trapped on cosmological scales is, say, $10\%$ of the 
equipartition energy, we expect that the gamma ray fluxes may be detectable
by GLAST for energies above 100 MeV, and by future ground based gamma ray 
telescopes such as VERITAS, MAGIC and HESS at higher energies. Even current 
observations with STACEE and HEGRA could actually provide interesting 
information about the abundance of cosmic rays in the intracluster gas of 
nearby clusters, as shown in fig. 1. 
Unfortunately, we are not aware of any scientific report 
of such attempt to look for high energy gamma rays with either STACEE or 
HEGRA. While the higher energy gamma ray emission is likely to be generated by
hadronic interactions, the lower energy gamma rays (in the MeV-GeV range)
can be generated by several processes related to electrons, and most of the
fluxes derived in the literature are in the range of interest for GLAST.

The importance of gamma ray observations can be appreciated particularly well
in the context of the growing multifrequency observations, that one piece at
a time, are allowing us to understand the processes that occur in the
intracluster volume, enriching it with hot gas, nonthermal particles and
magnetic fields, in a way that at present is still unclear.

\end{document}